\documentclass[nofootinbib,aps,twocolumn,showpacs,amsmath,amssymb,unsortedaddress]{revtex4}
\usepackage{amssymb}
\usepackage{amsmath}
\usepackage{epsfig}
\usepackage{amsfonts}
\usepackage{color}
\usepackage{float}
\usepackage{latexsym}
\usepackage{mathrsfs}
\usepackage{feynmf}
\usepackage{balance}
\usepackage{multirow}
\begin{document}
\title{ Local randomness in Cabello's non-locality argument from the Information Causality principle}
\author{Golnaz Zoka}
\affiliation{Department of Physics, Khayyam University,  Mashhad, Iran.}
\author{Ali Ahanj}
\email{a.ahanj@khayyam.ac.ir ; ahanj@ipm.ir}
\affiliation{Department of Physics, Khayyam University, Mashhad,Iran.}
\affiliation{Foundations of Physics Group, School of Physics,Institute for Research in Fundamental Science(IPM), P. O. Box 19395-5531, Tehran, Iran.}

%%%%%%%%%%%%%%%%%%%%%%%%%%%%%%%%%%%%%%%%%%%%%%%%%%%%%%%%%%%%%%%%%%%%%%%%%%%%%%%%%%%%%%%%
\begin{abstract}
The principle of non-violation of ``information causality", has been proposed as one of the foundational properties of nature\cite{nature}. The main goal of the paper is to explore the gap between quantum mechanical correlations and those allowed by ``information causality" in the context of local randomness by using Cabello's nonlocality argument.  This is interesting because the gap is slightly different than in the context of Hardy's similar nonlocality argument\cite{gazi}.
\end{abstract}
%%%%%%%%%%%%%%%%%%%%%%%%%%%%%%%%%%%%%%%%%%%%%%%%%%%%%%%%%%%%%%%%%%%%%%%%%%%%%%%%%%%%
\maketitle
 \section{Introduction}
One of the most amazing  features of Quantum Mechanics (QM) is the non-locality of correlations obtained by measuring entangled states. The nonlocal correlations  are neither caused by the exchange of a signal, nor are they due to pre-determined agreement as they violate Bell's inequalities \cite{bell,sca06}. On the other hand, Hardy \cite{har92,har93} and later Cabello \cite{cab02,lin,choud} gave a non-locality theorem to prove Bell's theorem without inequality. Notice that   Hardy nonlocality argument (HNA) is a special case of Cabello nonlocality argument CNA) but the important difference between them is that a mixed two-qubit entangled state can never exhibit HNA, but they can exhibit CNA \cite{kar}.\\
On the other hand, Popescu and Rohrlich \cite{pr} showed  that there exist non-signaling correlations that are more nonlocal than the ones predicted by QM  but still satisfy the no-signaling (NS) principle ( they called PR-box or NS-box).\cite{masane06}. The study of nonlocality  in the general no-signaling (NS) theorems (GNT) leads us toward a deeper understanding why such stronger than quantum correlations have not been observed in nature \cite{wang}. Also, the other important reason for studying GNT is that, it could be possible, QM is not the definitive description of nature, and more general correlations might be observed in the future. Thus, some physically motivated sets of correlations which are known to be bigger than the set of quantum correlations appear as candidates for possible generalizations of QM. In the first step, Van Dam showed that distant parties having access to the PR-box correlation can render communication complexity trivial and then argued that this could be a reason for the non-existence of such stronger nonlocal correlations in nature \cite{van}; another progress in this area was introduced in \cite{brassard}.\\
It has been shown that, some general physical principles  like Information Causality (IC)\cite{nature}, Macroscopic Locality (ML)\cite{nav09}, Exclusivity principle\cite{exp}, Relativistic Causality in the Classical limit \cite{roh1,roh2}, the uncertainty principle along with steering \cite{unp} and complementarity principle \cite{comp} are novel proposals to single out the quantum correlations, from rest of the no-signaling correlations, when two distant observers are involved \cite{das13}. Interestingly, most of these principles like IC and ML can explain the Tsirelson's bound of Bell Inequality in QM \cite{tsi80}.\\
 We also know that in GNT, maximum success probability for both HNA and CNA is $0.5$ \cite{ahanj,all09} while in QM the maximum probability of success for Cabello's case is not same as the Hardy's case \cite{har93,kunkri}( the maximum probability of success of HNA for two qubit systems is $0.09$, whereas in case of CNA it is $0.1078$ ). It has shown that by applying the IC principle \cite{nature} or the ML condition \cite{das13}, the maximum  success probability for both  HNA and CNA reduce to $0.20717$ (IC case) and $0.2063$ (ML case), but could not reach their respective quantum mechanical bounds \cite{ahanj,all09}. So, it remains this interesting research area, that whether some stronger necessary condition imposed by QM that are not reproduced from the IC and ML conditions. Recently, the authors in ref. \cite{gazi} shown that not only in terms of the value of the maximal probability of success, but also in terms of local randomness of the observable for Hardy's correlation, there is a gap between QM and IC condition. In this paper, we show that gap between QM and IC condition in context of Cabello's argument is much larger than the Hardy’s case.\\
 This article is organized as follows. In Sec.II, we introduce the all set of non-signaling correlations with binary input-output for two parties and next we study the set of non-signaling correlation that describe Cabello correlations. Next, we study CNA for two qubit system in the context of nonviolation of IC  and  local randomness (sec III). In sec.IV, we investigate about CNA in the context of local randomness in QM, and finally we bring our conclusions in sec V.
\section{The CNA under the NS-box}
 The all set of bipartite no-signaling correlations with binary input and output for each party is a convex set in a $2^4$ dimensional vector space. Let $P_{ab|XY}$ denote the joint probabilities, where $a,b \in \{0,1\}$  and
 $ X, Y\in\{0, 1\}$ denote output and inputs of parties, respectively. Then we have a set of $16$ joint probabilities  can be represented by a $4\times4$ correlation matrix:
    \begin{equation}\label{matrx}
\left(
  \begin{array}{cccc}
    P_{00|00} & P_{01|00} & P_{10|00} & P_{11|00} \\
    P_{00|01} & P_{01|01} & P_{10|01} & P_{11|01} \\
    P_{00|10} & P_{01|10} & P_{10|10} & P_{11|10} \\
    P_{00|11} & P_{01|11} & P_{10|11} & P_{11|11} \\
  \end{array}
\right)
\end{equation}
The joint probabilities satisfy:~i- Positivity:$ P_{ab|XY}\geq 0 ~ \forall a,b,X,Y \in \{0,1\} $,~ii- Normalization:~$\sum_{a,b=0}^{1} P_{ab|XY}=1 ~\forall X,Y \in \{0,1\}$~ iii- Marginal probabilities:~$P_{a|X}=\sum_{b} P_{ab|XY}=\sum_{b} P_{ab|X\bar{Y}}$ and $P_{b|Y}=\sum_{a} P_{ab|XY}=\sum_{a} P_{ab|\bar{X}Y}$. Notice that~$\bar{\alpha}=1\oplus \alpha$, where $\oplus$ is addition modulo $2$ .\\
Also, In the case of  binary inputs-outputs,  we can write \cite{all09}:
\begin{equation}\label{c_XY}
P_{ab|XY}=\frac{1}{4}[1+(-1)^a C_X +(-1)^b C_Y +(-1)^{a \oplus b}C_{XY}]
\end{equation}
the correlators $C_{XY}$ and marginal $C_X$ and $C_Y$ are defined by:
\begin{eqnarray}
  C_{XY} &=& P_{a=b|XY}-P_{a\neq b|XY}, \\
  C_X&=& P_{a=0|X}-P_{a=1|X}, \\
  C_Y &=& P_{b=0|Y}-P_{b=1|Y}.
\end{eqnarray}
The full set non-signaling correlations form an eight-dimensional polytope $\mathcal{P}$\cite{bar05}, where ~$24$ vertices of the polytope, $16$ of which represent local correlations (local vertices) and $8$ represent nonlocal correlation (nonlocal vertices). The local vertices can be expressed as:
\begin{equation}\label{local}
  P^{\alpha\beta\gamma\delta}_{ab|XY}=\delta_{(a=\alpha X \oplus \beta)}~\delta_{(b=\gamma Y \oplus \delta)}
\end{equation}
and the eight nonlocal vertices have the form:
\begin{equation}\label{nonlocal}
  P^{\alpha\beta\gamma}_{ab|XY}=\frac{1}{2}\delta_{a\oplus b= XY\oplus \alpha X \oplus \beta Y \oplus \gamma}
\end{equation}
%\end{document}
where $\alpha,\beta,\gamma,\delta\in \{0,1\}$ and notice that the canonical PR box corresponds to $PR box=P_{NL}^{000}$.\\
The set of local boxes forms a subpolytope of the full non-signalling polytope, and has facets which correspond to Bell inequalities:
\begin{equation}\label{bell}
B_{XY}=|\sum_{X^{\prime} ,Y^{\prime}}C_{X^{\prime},Y^{\prime}}-2C_{X,Y}|.
\end{equation}
According to the CHSH scenario, a mixture of local deterministic boxes must satisfy $B_{XY} \leq 2 $, and for an extreme non-local box, we have $B_{XY}=4$. The set of quantum boxes,($ B_{XY}\leq 2\sqrt{2}$) form a body with a smooth convex curve as its boundary.
% For binary inputs and outputs, Tsirelson, Landau and Masanes (TLM)\cite{tsi80,land, masane} have (independently) derived a necessary and sufficient criterion for a set of correlations $C_{XY}$ to admit a quantum description. In the form of Landau's criterion, a set of correlators $C_{XY}$  with unbiased marginal, must satisfy:
%\begin{equation}\label{landau}
%|C_{00}C_{10}-C_{01}C_{11}|\leq \sum_{j=0,1} \sqrt{(1-C_{0j}^2)(1-C_{1j}^2)}.
%\end{equation}
%Recently authors in \cite{navas07,navas08} proposed the criteria on quantum obtained correlators $C_{XY}$ with biased marginal.
%\section{Hardy-Cabello Nonlocality under no-signalling condition}
Now, let us consider joint probabilities satisfying the following constraints:
\begin{eqnarray}
  P_{ab|XY} &=& q_{1}, \nonumber\\
  P_{a^{\prime} \bar{b}|\bar{X} Y} &=& 0, \nonumber\\
  P_{\bar{a}b^{\prime}|X\bar{Y}} &=& 0, \nonumber\\
  P_{a^{\prime} b^{\prime}|\bar{X}\bar{Y}} &=& q_{4},
\end{eqnarray}
where $ \bar{X}$ ($\bar{Y}$) denotes complement of $X(Y)$ and $ a, a^{\prime}, b, b^{\prime}\in \{0,1\}$. These equations form the basis of Cabello nonlocality argument. It can easily be shown that these equations contradict local realism if $q_1 < q_4$. Whenever $q_1=0$, the Cabello's argument reduces to Hardy's argument. In the remaining part of this paper, without loss of generality we consider the following form of Cabello correlation:
$P_{11|11}=q_{4}>0 , P_{11|01}=q_{2}=0 ,  P_{11|10}=q_{3}=0 , P_{00|00}=q_{1} $. If we consider $q_1=0$ (Hardy's argument), then the above conditions, which can be written as a convex combination of the $ 5 $ of $16$ local vertices and one of the 8 nonlocal vertices:
\begin{eqnarray}\label{hardy}
  P^{H}_{ab|XY}&=&c_1 P^{0001}_{ab|XY}+c_2 P^{0011}_{ab|XY}+c_3 P^{0100}_{ab|XY}+c_4 P^{1100}_{ab|XY}\nonumber\\&+&c_5 P^{1111}_{ab|XY}+c_6 P^{001}_{ab|XY}
\end{eqnarray}
where $\sum^{6}_{i=1} c_i=1$. The other vertices can be covered by another set of Hardy's equations.The Cabello's non-locality argument ($q_1\neq 0$ and $q_1< q_4$, which can be written as a convex combination of the above $ 6 $ vertices which satisfies Hardy's conditions along with another four local vertices $P^{0000}_{ab|XY}$,~$P^{0010}_{ab|XY}$,~$P^{1000}_{ab|XY}$,~$P^{1010}_{ab|XY}$ and one nonlocal vertex $P^{110}_{ab|XY}$. So we get:
 \begin{eqnarray}\label{cabelo}
  P^{C}_{ab|XY}\nonumber&=&P^{H}_{ab|XY}+c_7 P^{0000}_{ab|XY}+c_8 P^{0010}_{ab|XY}\\&+&c_9 P^{1000}_{ab|XY}+c_{10} P^{1010}_{ab|XY}+c_{11} P^{110}_{ab|XY}
\end{eqnarray}
where the expression $P^{H}_{ab|XY}$ is given in Eq( \ref{hardy}) and coefficients $c_i$'s satisfy $\sum_{i=1}^{11}c_i=1$ . From here the correlation matrix for these Cabello's non-signaling boxes can be written as:
\begin{equation*}
  \left(
    \begin{array}{llll}
      c_7+c_8+c_9+c_{10} & c_1+c_2 & c_3+c_4&c_{5}\\
      c_2+c_7+c_9 & c_1+c_8+c_{10} & c_3+c_4+c_5& 0\\
      c_4+c_7+c_8 & c_1+c_2+c_5 & c_3+c_9+c_{10}& 0\\
      c_2+c_4+c_5+c_7&c_1+c_{8}&c_{3}+c_9&c_{10} \\
    \end{array}
  \right)
\end{equation*}

\begin{equation}\label{matrix2}
  +\frac{1}{2}\left(
    \begin{array}{llll}
      c_{11} &{c_6} &c_6&c_{11}\\
      0 & c_6+c_{11} & {c_6+c_{11}}& 0\\
      0 & c_6+c_{11} &c_6+c_{11}& 0\\
      {c_6}&c_{11}&c_{11}&{c_6} \\
    \end{array}
  \right)
\end{equation}\\
It is easy shown that the success probability for HNA is given by $q_2=P^H_{11|11}=\frac{c_6}{2}$. The no-signaling constraint imposes the maximum of $q_2$ is $0.5$ when $c_6=1$ and $c_1=c_2=c_3=c_4=c_5=0$. Similarly the success probability for CNA can be written as $P^c_{11|11}-P^c_{00|00}=(\frac{1}{2}c_6+c_{10})-(c_7+c_8+c_9+c_{10}+\frac{1}{2}c_{11})$, and here, too, we obtain that $(P^c_{11|11}-P^c_{00|00})_{max}=0.5$ for $c_6=1$ and rest of the $c_i's=0$ \cite{choud,jose}.
%\section{Cabello's Nonlocality under Information Causality}
\section{ CNA under IC in the context of local randomness (LR) }
In this section, we briefly review the principle of Information Causality (IC). Suppose that Alice and Bob, who are separated in space, have access to non-signaling resources such as shared randomness, entanglement or PR boxes. Alice receives a randomly generated N-bit string $\vec{X}=(X_0,X_1,...,X_{n-1})$ while Bob is asked to guess Alice's $i$-th bit where $i$ is randomly chosen from the set $\vec{Y}=\{0,1,2,...,N-1\}$ and $(M < N)$ classical bit are seat by Alice. The information causality principle impose that the total potential information \cite{nature} about Alice's bit string $\vec{X}$ accessible to Bob cannot exceed the volume of message the received from Alice, i.e.,
 \begin{equation}\label{ic}
   I= \sum_{i=1}^{N} I(X_i:\beta_i) \leq M,
 \end{equation}
 where  $I(X_i:\beta_i)$ is the Shannon mutual  information between the variable $X_i$ of Alice  and the Bob's guess $\beta$ . It was shown that \cite{nature} both classical and quantum correlations  satisfy the IC condition.
 Van Dam \cite{vandam} and Wolf and Wullschleger \cite{wull} by using a protocol,  derive a necessary condition for respecting the IC principle. The mathematical form of necessary conditions from Alice to Bob $(A \rightarrow B)$ can be expressed as:
 \begin{equation}\label{a-b}
   E_I^2+E^2_{II} \leq 1
 \end{equation}
where $E_i=2P_i^A-1$ for $i=\{I,II\}$ and
\begin{eqnarray}
  P_I^A &\equiv & \frac{1}{2}[P(a \oplus b=0|00)+P(a \oplus b=0|10)]\nonumber \\
   &=& \frac{1}{2}[P_{00|00}+P_{11|00}+P_{00|10}+P_{11|10}]
   \end{eqnarray}
   \begin{eqnarray}
    P_{II}^A &\equiv & \frac{1}{2}[P(a \oplus b=0|01)+P(a \oplus b=1|11)]\nonumber \\
   &=& \frac{1}{2}[P_{00|01}+P_{11|01}+P_{01|11}+P_{10|11}].
\end{eqnarray}
Also the necessary conditions for IC from Bob to Alice $(B\rightarrow A)$ can be expressed as:
\begin{equation}\label{b-a}
   F_I^2+F^2_{II} \leq 1,
 \end{equation}
where $F_i=2P_i^B-1$ for $i=\{I,II\}$ and
\begin{eqnarray}
  P_I^B &\equiv & \frac{1}{2}[P(a \oplus b=0|00)+P(a \oplus b=0|01)]\nonumber \\
   &=& \frac{1}{2}[P_{00|00}+P_{11|00}+P_{00|01}+P_{11|01}]
   \end{eqnarray}
   \begin{eqnarray}
    P^B_{II} &\equiv & \frac{1}{2}[P(a \oplus b=0|10)+P(a \oplus b=1|11)]\nonumber \\
   &=& \frac{1}{2}[P_{00|10}+P_{11|10}+P_{01|11}+P_{10|11}].
\end{eqnarray}
%\end{document}
It is important to note that the conditions $(E_I^2 +E_{II}^2 \leq 1)$ and $(F_I^2+F_{II}^2 \leq 1)$ are only a sufficient condition\cite{nature} for non-violating the IC principle. So a violation of ({\ref{a-b}}) or ({\ref{b-a}}) implies a violation of IC but the converse may not be true\cite{all09}. Now, it is easy to show that restricting Cabello's correlation (matrix (\ref{matrix2}) ) by imposing IC conditions (\ref{a-b}) and (\ref{b-a}), we get:\begin{eqnarray}
      (r-s)^2+(u-v)^2  &\leq & 1 \label{xx1}\\
      (u-s)^2+(r-v)^2 & \leq & 1 \label{xx2},
    \end{eqnarray}
    where $ r \equiv c_{8}-c_{2}$ , $s \equiv c_{1}+c_{3}+c_{6}-c_{7}$ ,
     $u\equiv c_{9}-c_{4}$ and $ v\equiv c_{5}+c_{6}+c_{10}.$\\
%\section{ Cabello's non-locality under Information Causality in the context of local randomness }
For the most general bipartite correlation an input $X$ on Alice's side is locally random if the marginal probabilities of all possible outcomes on Alice's side for this input, are equal and similarly for Bob \cite{gazi}. In other words, an input $X$ on Alice's side is locally random if for any choice of Bob's input$Y$ , we have:
\begin{eqnarray}\label{random1}
  P_{a=0|X}&=& \sum_b P_{0b|XY}=\frac{1}{2}~(\mathrm{the~case ~0_A})\\
  P_{a=1|X}&=& \sum_b P_{1b|XY}=\frac{1}{2} ~(\mathrm{the~ case~1_A}).
\end{eqnarray}
We show  these cases with $0_A$ and $1_A$ respectively. By using the Cabello's correlation matrix (\ref{matrix2}) ,we can find the conditions which coefficients $c_i$s must satisfy for the corresponding input to be locally random. For the case $0_A$, these conditions are:
\begin{eqnarray}\label{0a1a}
  c_1&+&c_2+c_7+c_8+c_9+c_{10}=\eta \\
  c_3&+&c_4+c_5=\eta
\end{eqnarray}
where $\eta \equiv \frac{1-c_6-c_{11}}{2}$ and for the case $1_A$, we have:
\begin{eqnarray}
  c_1&+&c_2+c_4+c_5+c_7+c_{8}=\eta \\
  c_3&+&c_9+c_{10}=\eta.
\end{eqnarray}
Similarly, an input $Y$ on Bob's side is locally random  if for any choice of Alice's input $X$ , we have:
\begin{eqnarray}
  P_{b=0|Y} &=& \sum_a P_{a0|XY}=\frac{1}{2}~\mathrm{(the~ case~ 0_B)} \\
  P_{b=1|Y} &=& \sum_a P_{a1|XY}=\frac{1}{2}~ \mathrm{(the~ case~ 1_B)}
\end{eqnarray}
%\end{document}
From the Cabello's correlation matrix (\ref{matrix2}), we can find the conditions which coefficients $c_i$s must satisfy for the corresponding input $0_B$ to be locally random:
\begin{eqnarray}\label{0a1a}
  c_3&+&c_4+c_7+c_8+c_9+c_{10}=\eta \\
  c_1&+&c_2+c_5=\eta
\end{eqnarray}
and for the case $1_B$, we have:
\begin{eqnarray}
  c_2&+&c_3+c_4+c_5+c_7+c_{9}=\eta \\
  c_1&+&c_8+c_{10}=\eta.
\end{eqnarray}
Now we try to see that for the no-signaling bipartite Cabello's correlations with two dichotomies observable on either side, what choices of inputs from  the set $\{0_A,1_A,0_B,1_B\}$ can be locally random. By using property of local randomness, for all possible choices of inputs that can be locally random for CNA, we can find the relation between $c_i$s for the corresponding choice of input to be locally random. We give the results for every case, in the TABLE I.\\
 By applying the IC conditions (Eqs (\ref{xx1}),(\ref{xx2})) for all possible choices of inputs that can be locally random for CNA and by using  MATLAB software, we find that in the all cases ( because cases 1-15 satisfy local randomness), always exist some $c_i$'s such that satisfying  the  two $IC$ inequalities, (Eqs (\ref{a-b}),(\ref{b-a}))~simultaneously (please see TABLE II). Thus we can conclude that for the cases $1-15$ , IC is satisfied, hence they can be true in quantum mechanics also. This is interesting results because in the Hardy case only the cases 9-15 satisfy $IC$ condition \cite{gazi}. Now we shall study cases 1-15 in the context of quantum mechanics.
\section{ CNA  under QM in the context of LR}
It is to be mentioned that, CNA  applies to both pure and mixed two-qubit entangled state, contrary to HNA runs  only for pure entangled state \cite{kar}. However, in this paper, we consider binary qubit system in the pure partial entangled state. From Schmidt decomposition, any pure entangled state of two particles can be written:
\begin{equation}\label{partial}
  |\psi>=\cos\beta|0>_A|0>_B+e^{i\gamma}\sin\beta|1>_A|1>_B,
\end{equation}
and the density matrix $\rho_{AB}$  can be written in terms of Pauli matrices as:
\begin{eqnarray}\label{density}
  &\rho_{AB}&=\frac{1}{4}\{\hat{I}^A \otimes \hat{I}^B +\cos2 \beta(\hat{I}^A\otimes \hat{Z}^B+\hat{Z}^A\otimes \hat{I}^B)\nonumber\\&+&\sin 2\beta \cos\gamma(\hat{X}^A\otimes \hat{X}^B- \hat{Y}^A\otimes \hat{Y}^B)\nonumber\\&+&\sin 2\beta \sin \gamma(\hat{X}^A\otimes \hat{Y}^B+ \hat{Y}^A\otimes \hat{X}^B)\nonumber\\&+&\hat{Z}^A\otimes \hat{Z}^B\}
\end{eqnarray}
where $\hat{X}\equiv\hat{\sigma}_x$, $\hat{Y}\equiv \hat{\sigma}_y $ and $\hat{Z}\equiv\hat{\sigma}_z$. The reduced density matrices $\rho_{A(B)}$  are:
\begin{equation}\label{roab}
  \rho_{A(B)}=tr_{B(A)}(\rho_{AB})=\frac{1}{2}\{\hat{I}^{A(B)}+\cos 2 \beta ~\hat{Z}^{A(B)}\}.
\end{equation}
Observable on Alice's (Bob's) side to be locally random, if:
\begin{equation}\label{localrandom}
Tr(\rho_{A(B)}P^{+})=Tr(\rho_{A(B)}P^{-}),
\end{equation}
where $P^{\pm}=\frac{1}{2}[I\pm \hat{n}.\sigma]$ are the projectors on the eigenstates of the observable on a single qubit $\hat{n}.\sigma$ and $\hat{n}=(\sin \theta \cos \phi, \sin \theta \sin \phi, \cos \theta)$ is any unit vectors in $\texttt{R}^3$.\\ For a partial entangled state an observable is locally random if and only if $\theta=\frac{\pi}{2}$ i.e. $\hat{n}=(\cos\phi, \sin\phi ,0)$. It is better to mention that for a maximally entangled state any arbitrary observable shows the property of local randomness \cite{gazi}, but Cabello's argument (like the Hardy case) doesn't run for a maximally entangled state.
%Now suppose A and $ A^{\prime}$ corresponds to $X=(0_A)$ and $ X=(1_A)$ are the observable on Alice's side and  B and $B^{\prime}$ corresponds to $ Y=(0_B)$ and $ Y=(1_B)$ are the observable on Bob's side and $a,b= 0(1) \longleftrightarrow +1(-1)$.
 The Cabello's correlation can be written as:

\begin{eqnarray}
P_{ab|XY}&=&\frac{1}{4}[\cos^2\beta(1+\tilde{a}\cos\theta_X)(1+\tilde{b}\cos\theta_Y)\nonumber\\&&+
\sin^2\beta(1-\tilde{a}\cos\theta_X)(1-\tilde{b}\cos\theta_Y)\nonumber\\ &&+ \sin2\beta \sin\theta_X\sin\theta_Y \cos(\varphi_X+\varphi_Y-\gamma)]\nonumber\\
\end{eqnarray}
 where $\tilde{a}=(-1)^a,\tilde{b}=(-1)^b \in \{1,-1\}$ and
 \begin{eqnarray*}
0_A&=&(\sin\theta_{X=0}\cos\theta_{X=0} \cos\varphi_{X=0}, \sin\theta_{X=0} \sin\varphi_{X=0},\\&& \cos\theta_{X=0})\\
1_A&=&(\sin\theta_{X=1}\cos\theta_{X=1} \cos\varphi_{X=1}, \sin\theta_{X=1} \sin\varphi_{X=1},\\&& \cos\theta_{X=1})
\end{eqnarray*}
are  observers  on Alice's side and
 \begin{eqnarray*}
0_B&=&(\sin\theta_{Y=0}\cos\theta_{Y=0} \cos\varphi_{Y=0}, \sin\theta_{Y=0} \sin\varphi_{Y=0},\\&& \cos\theta_{Y=0})\\
1_B&=&(\sin\theta_{Y=1}\cos\theta_{Y=1} \cos\varphi_{Y=1}, \sin\theta_{Y=1} \sin\varphi_{Y=1},\\&& \cos\theta_{Y=1})
\end{eqnarray*}
are observers on Bob's side.
From $q_2=P_{11|01}=0 $ and $q_3=P_{11|10}=0$, we obtain:
\begin{eqnarray}\label{mohem}
\cos^2\frac{\theta_{X=1}}{2}&=&( 1+ \tan^2 \beta \cot^2\frac{\theta_{Y=0}}{2})^{-1}\nonumber\\
\cos^2\frac{\theta_{Y=1}}{2}&=&( 1+ \tan^2 \beta \cot^2\frac{\theta_{X=0}}{2})^{-1}.
\end{eqnarray}
 Finally by substituting Eq (\ref{mohem}) in $q_4-q_1=P_{11|11}-P_{00|00}$, we get:
\begin{eqnarray}\label{q4q1}
  &&q_4-q_1 = \frac{sin^2\beta(\tan \beta-\tan\frac{\theta_{X=0}}{2}\tan\frac{
\theta_{Y=0}}{2})^2}
{(\tan^2\frac{\theta_{X=0}}{2}+\tan^2 \beta)(\tan^2\frac{\theta_{Y=0}}{2}+\tan^2 \beta)} \nonumber\\
   &-& (\cos \beta \cos\frac{\theta_{X=0}}{2}\cos\frac{\theta_{Y=0}}{2}-\sin\beta \sin \frac{\theta_{X=0}}{2}\sin\frac{\theta_{Y=0}}{2})^2.\nonumber\\
\end{eqnarray}
Now by using the local randomness condition $(\theta=\frac{\pi}{2})$ and from equation (\ref{mohem}) we find that: i- the cases $1,2,3,4,5,10,11$ imply that $\beta=\frac{\pi}{4}$ and it implies that the state has to be a maximally entangled state and cannot be consider( TABLE III). ii- On the other hand, the cases 6, 7, 8 and 9 don't violate inequality (\ref{q4q1}) i.e. $(q_4-q_1)_{Max}=0$. iii- But for the case of just one observable to be locally random ( cases 12, 13, 14, 15), we find that there are non-maximally entangled states such that the maximum of $q_4-q_1$ larger than zero and therefore for these cases Cabello's argument runs. We can now conclude that for a quantum mechanical state showing Cabello's nonlocality, ( same Hardy case), at most one out of the four observable can be locally random (please refer to TABLE III).
\section{ Conclusion}
We know that the maximum success probability of HNA and CNA for the class of generalized no-signaling theories (GNT) in the binary qubit systems  is $0.5$ \cite{choud,jose}. The authors in ref.\cite{ahanj} showed that on applying the principle of information causality (IC) this bound decreases from $0.5$ to $0.20717$ in both cases but could not reach their respective quantum mechanical bound. It is interesting that in QM the maximum probability of success of HNA for two-qubit system is $0.09$, that is not same as the CNA, where it is $0.1078$ \cite{kunkri}. We study all the possibilities of local randomness in Cabello's correlation respected by the principle of non-violation of IC. We observe that not only in terms of the value of the maximal probability of success, but also in terms of local randomness there is a gap between QM and IC condition, and this gap is  larger than the Hardy's case \cite{gazi}. This is interesting because the gap is slightly different than in the context of Hardy's similar nonlocality argument. This difference may be relevant in assessing the viability of ``information causality" (and ``macroscopic locality") as partial candidate explanations for why QM correlations are weaker than generalized non-signalling correlations. However, it remains to see, in future, whether some stronger necessary condition for IC can close the gap between QM and IC. So, in this paper, we find that in the CNA case, the number of non-quantum correlation definitely obey the IC condition is more than the HNA case. Therefore,  we can conclude our work that the optimal success probability for CNA in QM is
stronger control than the HNA for detecting post-quantum no signaling correlations. Finally, it is better mention that, the IC condition (and also another principles ) is not sufficient for distinguishing quantum correlations from all post-quantum correlation which are bellow the Tsirelson's bound. So, it remains open question that, does the full power of IC ( some other conditions derived from IC) eliminate remaining post-quantum correlations obey the Tsirelson's bound?
\section{Acknowledgments}
This work is partially supported by Foundation of Physics Group,Institute for Research in Fundamental Science, Tehran, Iran.

\begin{table}
\caption{Relation between $c_i$s for the corresponding choice of inputs to be locally random($\eta=\frac{1-c_6-c_{11}}{2}$).}
\centering

\begin{center}
    \begin{tabular}{ | l | l  | p{5cm} |}

        \hline
    Case & Locally random inputs & Relation between $c_i$s   \\ \hline
    1 & $0_A,1_A,0_B,1_B$  & $c_2=c_4=c_7=c_8=c_9=0$, $c_1=c_3=\eta-c_5$, $c_5=c_{10} $\\ \hline \hline
    2& $0_A,1_A,0_B $ & $ c_4=c_7=c_8=0, c_3+c_5=\eta , c_1+c_2=c_3, c_9+c_{10}=c_5$\\ \hline
    3 & $0_A,1_A,1_B$ & $c_2=c_7=c_9=0, c_3+c_{10}=\eta , c_1+c_8=c_3, c_4+c_5=c_{10} $  \\
    \hline
    4& $0_A,0_B,1_B$ & $c_2=c_7=c_9=0, c_1+c_{5}=\eta , c_3+c_4=c_1, c_8+c_{10}=c_{5} $  \\\hline
    5& $1_A,0_B,1_B$ & $c_4=c_7=c_8=0, c_1+c_{10}=\eta , c_3+c_9=c_1, c_2+c_{5}=c_{10} $\\ \hline \hline
    6& $0_A,1_A$ & $c_3+c_4+c_5=\eta , c_3=c_1+c_{2}+c_7+c_8 , c_4+c_5=c_9+c_{10} $\\ \hline
    7& $0_B,1_B$ & $c_1+c_2+c_5=\eta , c_1=c_3+c_{4}+c_7+c_9 , c_2+c_5=c_8+c_{10} $\\ \hline
    8& $1_A,1_B$ & $c_1+c_8=c_3+c_9=\eta-c_{10} , c_{10}=c_2+c_{4}+c_5+c_7$\\ \hline
    9& $0_A,0_B$ & $c_1+c_2=c_3+c_4=\eta-c_{5} , c_{5}=c_7+c_{8}+c_9+c_{10}$\\ \hline
    10& $0_A,1_B$ & $c_2=c_7=c_9=0 ,c_3+c_4+c_5=c_1+c_8+c_{10}=\eta$\\ \hline
    11& $1_A,0_B$ & $c_4=c_7=c_8=0 ,c_1+c_2+c_5=c_3+c_9+c_{10}=\eta$\\ \hline \hline
    12& $0_A$& $c_1+c_2+c_7+c_8+c_9+c_{10}=c_3+c_4+c_5=\eta $\\ \hline
    13& $1_A$& $c_1+c_2+c_4+c_5+c_7+c_{8}=c_3+c_9+c_{10}=\eta $\\ \hline
    14& $0_B$& $c_3+c_4+c_7+c_8+c_9+c_{10}=c_1+c_2+c_5=\eta $\\ \hline
    15& $1_B$& $c_2+c_3+c_4+c_5+c_7+c_{9}=c_1+c_8+c_{10}=\eta $\\ \hline \hline
    \end{tabular}
\end{center}

\end{table}
%%%%%%%%%%%%%%%%%%%%%%%%%%%%%%%%%%%%
\begin{table}[hb]
\caption{Some of $c_i$'s can  satisfy both two inequalities IC condition, simultaneously .}
\centering

      \begin{center}
    \begin{tabular}{| c | l | c | c | c | c | c | c | c | c | c | c |c |c | p{2cm}|}

        \hline
   Case & Locally random inputs &$c_1$&$c_2$&$c_3$&$c_4$&$c_5$& $c_6$&$c_7$&$c_8$&$c_9$&$c_{10}$&$c_{11}$&$ E_I+E_{II}\leq 1$ & $F_I+F_{II}\leq 1 $\\ \hline

   \multirow{3}{*}{1} & \multirow{3}{*}{$ 0_A,1_A,0_B,1_B$}
    &0&0&0&0&0&0&0&0&0&0&1&0&0\\
    & &0&0&0&0&0.1000&0&0&0&0&0.1000&0.8000&0.0400&0.0400\\
    & &0&0&0&0&0.4000&0.1000&0&0&0&0.4000&0.1000&0.8200&0.8200\\ \hline\hline
\multirow{3}{*}{2} & \multirow{3}{*}{$ 0_A,1_A,0_B$}
 & 0 &0&0&0&0&0&0&0&0&0&1&0&0  \\
& &0&0&0&0&0.4000&0.2000&0&0&0.4000&0&0&0.0800&0.4000\\
& &0&0&0&0&0.5000&0&0&0&0.5000&0&0&0&0.5000\\ \hline
 \multirow{3}{*}{3} & \multirow{3}{*}{$ 0_A,1_A,1_B$}
  &0&0&0&0& 0 & 0 & 0 & 0 & 0 & 0 & 1 & 0 & 0  \\
  & &0&0&0&0&0&0.1000&0&0&0&0&0.9000&0.0200&0.0200\\
 & &0&0&0&0&0.4000&0&0&0&0&0.4000&0.2000&0.6400&0.6400\\ \hline
\multirow{3}{*}{4} & \multirow{3}{*}{$ 0_A,0_B,1_B$}
 & 0 & 0 & 0 & 0 & 0 & 0 & 0& 0 & 0 & 0 & 1 & 0 & 0  \\
& &0&0&0&0&0.2000&0.6000&0&0.2000&0&0&0&0.8000&0.7200\\
& &0&0&0&0&0.5000&0&0&0.5000&0&0&0&0.5000&0\\ \hline
 \multirow{3}{*}{5} & \multirow{3}{*}{$1_A,0_B,1_B$}
 & 0 & 0 & 0 & 0 & 0 & 0 & 0 & 0 & 0 & 0 & 1 & 0 & 0
  \\& &0&0&0&0&0.2000&0&0&0&0&0.2000&0.6000&0.1600&0.1600\\
 & &0&0&0&0&0.3000&0.3000&0&0&0&0.3000&0.1000&0.9000&0.9000\\ \hline\hline
\multirow{3}{*}{6} & \multirow{3}{*}{$ 0_A,1_A,$}
& 0 & 0 & 0 & 0 & 0 & 0 & 0& 0 & 0 & 0 & 1 & 0 & 0  \\
& &0&0&0&0&0.3000&0&0&0&0.3000&0&0.4000&0&0.1800\\
& &0&0&0&0&0.5000&0&0&0&0.5000&0&0&0&0.5000\\ \hline
 \multirow{3}{*}{7} & \multirow{3}{*}{$0_B,1_B$}
  & 0 & 0& 0 & 0 & 0 & 0 & 0 & 0 & 0 & 0 & 1 & 0 & 0  \\
  & &0&0&0&0&0.4000&0&0&0.4000&0&0&0.2000&0.3200&0\\
  & &0&0&0&0&0.5000&0&0&0.5000&0&0&0&0.5000&0\\ \hline
\multirow{3}{*}{8} & \multirow{3}{*}{$1_A,1_B$}
& 0 & 0 & 0 & 0 & 0 & 0& 0 & 0 & 0 & 0 & 1 & 0 & 0  \\
& &0&0&0&0&0&0&0&0.5000&0.5000&0&0&0.5000&0.5000\\
& &0&0&0&0&0.3000&0.3000&0&0&0&0.3000&0.1000&0.9000&0.9000\\ \hline
 \multirow{3}{*}{9} & \multirow{3}{*}{$ 0_A,0_B$}
  & 0 & 0 & 0 & 0 & 0 & 0 & 0 & 0 & 0 & 0 & 1 & 0 & 0 \\
  & &0&0&0&0&0.3000&0&0.3&0&0&0&0.4000&0.1800&0.1800\\
  & &0&0&0&0&0.5000&0&0.5&0.1000&0&0&0&0.5000&0.3200\\ \hline
\multirow{3}{*}{10} & \multirow{3}{*}{$ 0_A,1_B$}
& 0 & 0 & 0 & 0 & 0 & 0 & 0 & 0 & 0 & 0 & 1 & 0 & 0  \\
& &0&0&0&0&0.1000&0&0&0.1000&0&0&0.8000&0.0200&0\\
& &0&0&0&0&0.5000&0&0&0.3000&0&0.2000&0&0.5800&0.1600\\ \hline
 \multirow{3}{*}{11} & \multirow{3}{*}{$1_A,0_B$}
  & 0 & 0 & 0 & 0 & 0 & 0 & 0 & 0 & 0 & 0 & 1 & 0 & 0  \\
  & &0&0&0&0&0.3000&0&0&0&0.3000&0&0.4000&0&0.1800\\
  & &0&0&0&0&0.5000&0&0&0&0.5000&0&0&0&0.5000\\ \hline\hline
\multirow{3}{*}{12} & \multirow{3}{*}{$0_A$}
& 0 & 0 & 0 & 0 & 0 & 0 & 0 & 0 & 0 & 0 & 1 & 0 &0 \\
& &0&0&0&0&0.1000&0&0&0&0.1000&0&0.800&0&0.2000\\
& &0&0&0&0&0.5000&0&0&0&0.3000&0.2000&0&0.1600&0.5800\\ \hline
 \multirow{3}{*}{13} & \multirow{3}{*}{$1_A$}
  & 0 & 0 & 0 & 0 & 0 & 0 & 0 & 0 &0 & 0 & 1 & 0 &0  \\
  & &0&0&0&0&0&0&0&0.5000&0&0.5000&0&0.5000&0\\
  & &0&0&0&0&0.3000&0.4000&0&0&0.3000&0&0&0.3200&0.5000\\ \hline
\multirow{3}{*}{14} & \multirow{3}{*}{$0_B$}
 & 0 & 0 & 0 & 0 & 0 & 0 & 0 & 0 & 0 & 0 & 1 & 0 & 0  \\
& &0&0&0&0&0.1000&0&0&0&0.1000&0&0.8000&0&0.0200\\
& &0&0&0&0&0.5000&0&0.5&0&0&0&0&0.5000&0.5000\\ \hline
 \multirow{3}{*}{15} & \multirow{3}{*}{$1_B$}
 & 0 & 0 & 0 & 0 & 0 & 0 & 0 & 0 & 0 & 0 & 1 & 0 & 0  \\
 & &0&0&0&0&0&0&0.1&0.1000&0.4000&0.4000&0&0.4000&0.3400\\
 & &0&0&0&0&0.3000&0&0&0.4000&0.2000&0.1000&0&0.2000&0.0400\\ \hline
 \end{tabular}
\end{center}

\end{table}

\begin{table}
\caption{The Maximum value of $q_4-q_1$ for the corresponding choice of inputs to be locally random..}
\centering

\begin{center}
    \begin{tabular}{| l | l | c | c | c | c | c  | p{2cm}|}

        \hline
   Case & Locally random inputs & $\theta_A$ & $\theta_{A^{\prime}}$ &   $\theta_B$ & $\theta_{B^{\prime}}$ &  $\beta$ & $ (q_4-q_1)_{MAX}$\\ \hline

    1 & $0_A,1_A,0_B,1_B$  & $ \frac{\pi}{2}$ & $ \frac{\pi}{2}$ & $ \frac{\pi}{2}$ &$ \frac{\pi}{2}$ &$ \frac{\pi}{4}$ & 0 \\ \hline \hline
    2& $0_A,1_A,0_B $ &$ \frac{\pi}{2}$ & $ \frac{\pi}{2}$ & $ \frac{\pi}{2}$ &$ 0 ~to~\pi$ &$ \frac{\pi}{4}$ & 0 \\ \hline

    3 & $0_A,1_A,1_B$ & $ \frac{\pi}{2}$ & $ \frac{\pi}{2}$ & $ 0~to~\pi$ &$ \frac{\pi}{2}$ &$ \frac{\pi}{4}$ & 2.1570e-032  \\ \hline

    4& $0_A,0_B,1_B$ & $ \frac{\pi}{2}$ & $ 0~to~\pi$ & $ \frac{\pi}{2}$ &$ \frac{\pi}{2}$ &$ \frac{\pi}{4}$ & 0 \\ \hline

    5& $1_A,0_B,1_B$ & $ 0~to~\pi$ & $ \frac{\pi}{2}$ & $ \frac{\pi}{2}$ &$ \frac{\pi}{2}$ &$ \frac{\pi}{4}$ & $8.3267e-017$\\ \hline \hline

    6& $0_A,1_A$ & $ \frac{\pi}{2}$ & $ \frac{\pi}{2}$ & $ 0~to~\pi$ &$ 0~to~\pi$ &$ 0~to~\frac{\pi}{2}$ & $3.0815e-033$\\ \hline

    7& $0_B,1_B$ & $ 0~to~\pi$ & $ 0~to~\pi$ & $ \frac{\pi}{2}$ &$ \frac{\pi}{2}$ &$0 ~to~ \frac{\pi}{2}$ & $3.0815e-033$\\ \hline

    8& $1_A,1_B$ & $0 ~to~\pi$ & $\frac{\pi}{2}$ & $0~to~\pi$ & $\frac{\pi}{2} $ & $0~to~\frac{\pi}{2}$ & 0\\ \hline

    9& $0_A,0_B$ & $\frac{\pi}{2}$ & $0 ~to~\pi$ & $\frac{\pi}{2}$ & $0~to~\pi$ & $ 0 ~to~\frac{\pi}{2}$ & $2.1570e-032$\\ \hline

    10& $0_A,1_B$ &$ \frac{\pi}{2} $ & $ 0 ~to~\pi$ & $0~to~\pi$ & $\frac{\pi}{2} $ & $\frac{\pi}{4}$ & $1.1102e-016$ \\ \hline

    11& $1_A,0_B$ & $0 ~to~\pi$ & $\frac{\pi}{2}$ & $ \frac{\pi}{2} $ &  $0~to~\pi$ & $\frac{\pi}{4}$ & $ 8.3267e-017$\\ \hline \hline

    12& $0_A$& $\frac{\pi}{2}$ & $0 ~to~\pi$ & $0 ~to~\pi$ &$0 ~to~\pi$ &$0 ~to~\frac{\pi}{2}$ & $0.0990$\\ \hline

    13& $1_A$ &  $0 ~to~\pi$ &$\frac{\pi}{2}$ &$0 ~to~\pi$ &$0~to~\pi$ &$0~to~\frac{\pi}{2}$ &$0.0714$  \\ \hline

    14& $0_B$& $0 ~to~\pi$ &$0 ~to~\pi$ & $\frac{\pi}{2}$ &$0 ~to~\pi$ &$0 ~to~\frac{\pi}{2}$& $0.0990$ \\ \hline

    15& $1_B$& $0 ~to~\pi$ &$0 ~to~\pi$ &$0 ~to~\pi$ &$\frac{\pi}{2}$ &$0 ~to~\frac{\pi}{2}$ & $0.0714$\\ \hline \hline
    \end{tabular}
\end{center}

\end{table}

%%%%%%%%%%%%%%%%%%%%%%%%%%%%%%%%%%%%
%%%%%%%%%%%%%%%%%%%%%%%%%%%%%%%%%%%%%%%%

\end{document}